\documentclass[journal]{IEEEtran}

\usepackage[nolist]{acronym}
\usepackage{amsmath}
\usepackage[table,xcdraw]{xcolor}
\usepackage{balance}
 
\ifCLASSINFOpdf
\usepackage[pdftex]{graphicx}
\else
\fi
\begin{document}

\title{A Novel Airborne Self-organising Architecture for 5G+ Networks}

\author{\IEEEauthorblockN{Hamed Ahmadi,~\IEEEmembership{Senior Member,~IEEE,} Konstantinos Katzis,~\IEEEmembership{Senior Member,~IEEE,} and \\Muhammad Zeeshan Shakir,~\IEEEmembership{Senior Member,~IEEE}
\thanks{H. Ahmadi is with the School of Electrical and Electronic Engineering, University College Dublin, Ireland, Email: hamed.ahmadi@ucd.ie.}
\thanks{K. Katzis is with the Department of Computer Science and Engineering, European University Cyprus, Cyprus, Email: K.Katzis@euc.ac.cy.}
\thanks{M. Z. Shakir is with the School of Engineering and Computing, University of the West of Scotland, Paisley, Scotland, UK, Email: muhammad.shakir@uws.ac.uk.}
}

}

%\author{Michael~Shell,~\IEEEmembership{Member,~IEEE,}
%        John~Doe,~\IEEEmembership{Fellow,~OSA,}
%        and~Jane~Doe,~\IEEEmembership{Life~Fellow,~IEEE}}% <-this % stops a space
%\thanks{M. Shell was with the Department
%of Electrical and Computer Engineering, Georgia Institute of %Technology, Atlanta,

\maketitle

\begin{abstract}
\ac{NFPs} such as unmanned aerial vehicles, unmanned balloons or drones flying at low/medium/high altitude can be employed to enhance network coverage and capacity by deploying a swarm of flying platforms that implement novel radio resource management techniques. In this paper, we propose a novel layered architecture where \ac{NFPs}, of various types and flying at low/medium/high layers in a swarm of flying platforms, are considered as an integrated part of the future cellular networks to inject additional capacity and expand the coverage for exceptional scenarios (sports events, concerts, etc.) and hard-to-reach areas (rural or sparsely populated areas). Successful roll-out of the proposed architecture depends on several factors including, but are not limited to: network optimisation for NFP placement and association, safety operations of NFP for network/equipment security, and reliability for NFP transport and control/signaling mechanisms. In this work, we formulate the optimum placement of NFP at a \ac{LL} by exploiting the airborne \ac{SON} features. Our initial simulations show the NFP-LL can serve more \ac{UE}s using this placement technique.
\end{abstract}


%\vspace{-0.5in}
\section{Introduction}
%\ac{NFPs} such as unmanned aerial vehicles, unmanned balloons or drones flying at low/high/medium altitude can be employed to enhance network coverage and capacity by deploying a swarm of flying platforms that implement novel radio resource management techniques \cite{icc2017wk}. In this paper, we propose a new architecture where \ac{NFP}, of various types and flying at different altitudes in a swarm of flying platforms, are considered as an integrated part of the future cellular networks to inject additional capacity and expand the coverage for exceptional scenarios (sports events, concerts, etc) and hard-to-reach areas (rural or sparsely populated areas). Successful roll-out of the proposed architecture depends on several factors including, but not limited to: network optimisation for NFP placement and association, safety operation of NFP for network/equipment security, and reliability for NFP transport and control/signaling mechanisms. This work also presents an efficient placement mechanism that can be easily incorporated in the proposed architecture. Our results show that using this \ac{SON} features \ac{NFP} can serve more \ac{UE}s. 

The \ac{5G} networks, which are expected to
be rolled out soon after 2020, will support a 1000 times higher average data traffic, 10-100 fold increase in data rate and connected devices \cite{6815890}. One of the enabling solutions to meet the demand for high data rate is the densification in cellular network by complementing  the ultra-dense deployment of low power small base stations (SBSs) with the airborne cellular network and forming a multi-tier heterogeneous network (HetNet) for \ac{5G}+ systems.  

One of the biggest challenges in designing such an airborne cellular network communication system, is to optimally position the flying \ac{BS} or drone-cell and maintain that position so that the network can benefit the most. Extensive work by numerous research groups investigated how the position of the flying \ac{BS} or drone-cell can affect the performance of the aerial and terrestrial communication systems as well how its position can play a vital role in the operation of the network. For instance, in \cite{c2b90718cf1640aa91ce30953aae3b57} authors looked at the effect of different generic mobility models that can be used to characterize the pertinent movements of HAPs and their effect on the communications. Furthermore, system operating parameters such as the flight path, operation location and service duration over a particular service area can be of high significance in terms of optimized communications for the users on the ground if the aim is to provide seamless connectivity during handoff between platforms \cite{317e1315569e447aa94e59d77455937a}. \cite{ROHDE20131893} introduces a genetic Interference-aware Positioning of Aerial Relays (IPAR) algorithm which is able to find suitable positions for the UAVs that maximize the downlink throughput of the cellular network. Similarly, \cite{7122575} discusses efficient algorithms developed to compute the optimal position of the drone for maximizing the data rate, which are shown to be highly effective via simulations.  In \cite{6863654}, authors presents that the optimal altitude is a function of the maximum allowed pathloss and of the statistical parameters of the urban environment, as defined by the International Telecommunication Union. They also present a closed-form formula for predicting the probability of the geometrical line of sight between a \ac{LAP} and a ground receiver. In \cite{7510820} authors present a 3-D placement problem with the objective of maximizing the revenue, which is measured by the maximum number of users covered by the drone-cell. To solve this problem, they have formulated an equivalent problem which was solved efficiently to find the location and size of the coverage region, and the altitude of the drone-cell. 

\begin{figure*}[t]
	\begin{center}
		\includegraphics[width=0.9\textwidth]{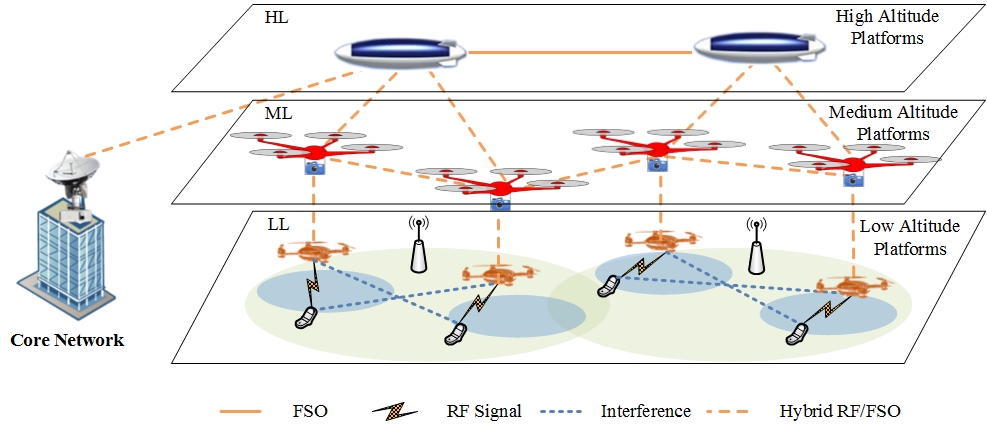}\\
	\end{center}
	\caption{Graphical illustration of the hierarchical airborne self organizing architecture where variety of \ac{NFPs} are flying at different altitudes and are offering a complementary wireless network.}
	\label{illustrative}
\end{figure*}

An airbone cellular network comprises of \ac{NFP}s of various types including  \ac{UAV}s, drones, balloons, and high-altitude/medium-altitude/low-altitude platforms (HAPs/MAPs/LAPs) to expand the cellular coverage and deliver Internet services to remote and dedicated regions where infrastructure is not available and expensive to deploy. However, a major challenge in such networks is how to design and manage Self-organisation in cellular networks which is mostly seen as a result of distributed decision making \cite{aliu2013_SONsurvey}. In conventional \ac{SON}, self-configuration, self-optimization and self-healing are considered as main \ac{SON} functions. An airborne cellular network has a more dynamic nature compared to a fixed cellular network because the position of its elements may change over the time. This may be due to change in demand, weather conditions, coverage requirements, battery limitations and even due to some real-time traffic changes/abnormalities in the network. In such an airborne cellular network each \ac{BS} i.e., \ac{NFP}-\ac{LL}, must be capable of performing a combination of conventional and emerging \ac{SON} functionalities for airborne cellular network depending on the role of \ac{NFP}s in the network. 
 
The rest of the paper is organised as follows: A novel airborne cellular architecture is introduced in Section II. Section III presents optimisation problem of the proposed airborne SON and its implementation. Section IV presents simulation results and some discussions on the results. Conclusions are drawn under section V and finally some challenges are summarised in Section VI. 

%\section{Review}

%http://www.absolute-project.eu/images/ABSOLUTE_white_paper_2015.pdf

%\subsection{special characteristics of surveillance}
%\subsection{Our contribution: layer-based new architecture for drone surveillance}

%\section{Our contribution: layer-based new architecture for drone surveillance}

%\section{system architecture}
\section{Airborne Self Organising Networks}

In an emerging airborne cellular network, the \ac{SON} functionalities can be realized differently. For example, when a cell outage is detected, neighboring \ac{BS}s in conveantional \ac{SON} compensate this outage by increasing their transmission power or changing their antenna tilt (self-healing) while in an airborne \ac{SON}, \ac{NFP}s can also effectively reorganise their positions to compensate the outage with lower energy consumption. The new placements can be decided centrally or locally at each \ac{NFP} element -- distributed decision making.

\subsection{Proposed Systems Architecture:}

Fig. 1 shows our proposed novel layered architecture where \ac{NFPs} are distributed in a hierarchal manner such that the \ac{LL} is responsible for access or fronthaul for small cells, \ac{HL} is responsible for transport network and \ac{ML} is serving as a relay between the two layers. \ac{NFPs} in the \ac{LL} are typical low altitude platforms (LAPs) flying at relatively lower altitudes and responsible for network optimisation including NPF placement and association based on resource allocation, interference management, etc. On the other side, \ac{NFPs} such as \ac{UAV}, unmanned balloons or drones flying at low/high/medium altitude can be used to enhance network coverage and capacity by deploying a swarm of flying platforms that implement novel radio resource management techniques \cite{icc2017wk}. In this paper,  operating in the \ac{HL} belong to \ac{HAP} category and are responsible for optimising the resources in transport networks for lower layers. \ac{NFPs} in the \ac{ML} belong to the \ac{MAP} category and are responsible for relaying the network between the lower and higher layers in our proposed architecture. Safety and security of the proposed \ac{NFPs} driven cellular network is of significant importance as such network deployment are more prone to the associated network security risks including their safe operation and security of information. In our architecture, \ac{NFPs} in the medium layer are dual role playing i.e., in addition to relaying, medium altitude platforms (MAPs) are performing surveillance to ensure safe and secure operation of the architectures. The surveillance operation includes network monitoring, surveillance scheduling, and decisions to further optimise the network and ensure reliable and secure operation by exploiting the collaboration between the participating \ac{NFPs} in the medium layer. \ac{NFPs} in the lower layer are optimally distributed to offer capacity and expand coverage via resource and interference management. It is to note that in the proposed architecture HAPs and MAPs are flying with fixed/known locations, however, LAPs are flying with relatively random or optimally distributed locations and offering coverage or capacity in an optimal way. 

\begin{table*}[t!]
\centering
\caption{Summary of aerial platforms participaiting in Airbone cellular network.}
\label{my-label}
\resizebox{\textwidth}{!}{
\begin{tabular}{lccccccccc}
\multicolumn{1}{c}{\cellcolor[HTML]{C0C0C0}\textbf{Platform Name}} & \cellcolor[HTML]{C0C0C0}\textbf{\begin{tabular}[c]{@{}c@{}@{}c@{}}AirShip\\AirPlane\\AirBaloon\\AirCopter\end{tabular}} & \cellcolor[HTML]{C0C0C0}\textbf{\begin{tabular}[c]{@{}c@{}}Manned\\Unmanned\end{tabular}} & \cellcolor[HTML]{C0C0C0}\textbf{\begin{tabular}[c]{@{}c@{}}Max  Altitude \\ (approximately)\end{tabular}} & \cellcolor[HTML]{C0C0C0}\textbf{\begin{tabular}[c]{@{}c@{}}Platform\\length\end{tabular}} & \cellcolor[HTML]{C0C0C0}\textbf{\begin{tabular}[c]{@{}c@{}}Platform width\\ (Wing Span)\end{tabular}} & \cellcolor[HTML]{C0C0C0}\textbf{\begin{tabular}[c]{@{}c@{}}Platform\\weight\end{tabular}} & \cellcolor[HTML]{C0C0C0}\textbf{Range} & \cellcolor[HTML]{C0C0C0}\textbf{\begin{tabular}[c]{@{}c@{}}Max\\Payload\end{tabular}} & \cellcolor[HTML]{C0C0C0}\textbf{Endurance}\\
\cellcolor[HTML]{C0C0C0}\textbf{} & \cellcolor[HTML]{C0C0C0}\textbf{(S/P/B/C)} & \cellcolor[HTML]{C0C0C0}\textbf{(M/U)} & \cellcolor[HTML]{C0C0C0}\textbf{(m)} & \cellcolor[HTML]{C0C0C0}\textbf{(m)} & \cellcolor[HTML]{C0C0C0}\textbf{(m)} & \cellcolor[HTML]{C0C0C0}\textbf{(kg)} & \cellcolor[HTML]{C0C0C0}\textbf{(km)} & \cellcolor[HTML]{C0C0C0}\textbf{(kg)} & \cellcolor[HTML]{C0C0C0}\textbf{(hrs)}\\
\multicolumn{1}{l}{} & \multicolumn{1}{l}{} & \multicolumn{1}{l}{} & \multicolumn{1}{l}{} & \multicolumn{1}{l}{} & \multicolumn{1}{l}{} & \multicolumn{1}{l}{} & \multicolumn{1}{l}{} & \multicolumn{1}{l}{} \\ 
\cellcolor[HTML]{C0C0C0}\textbf{Lower Layer  - LAP} & \cellcolor[HTML]{C0C0C0}\textbf{} & \cellcolor[HTML]{C0C0C0}\textbf{} & \cellcolor[HTML]{C0C0C0}\textbf{} & \cellcolor[HTML]{C0C0C0}\textbf{} & \cellcolor[HTML]{C0C0C0}\textbf{} & \cellcolor[HTML]{C0C0C0}\textbf{} & \cellcolor[HTML]{C0C0C0}\textbf{} & \cellcolor[HTML]{C0C0C0}\textbf{} & \cellcolor[HTML]{C0C0C0}\textbf{} \\
\cellcolor[HTML]{FFFFFF}Amazon Drone & \cellcolor[HTML]{FFFFFF}C & \cellcolor[HTML]{FFFFFF}U & \cellcolor[HTML]{FFFFFF}122 & \cellcolor[HTML]{FFFFFF} & \cellcolor[HTML]{FFFFFF} & \cellcolor[HTML]{FFFFFF}25 & \cellcolor[HTML]{FFFFFF}16 & \cellcolor[HTML]{FFFFFF}2.26 & \cellcolor[HTML]{FFFFFF}0.5 \\
\cellcolor[HTML]{FFFFFF}Aerovironment Dragon Eye & \cellcolor[HTML]{FFFFFF}P & \cellcolor[HTML]{FFFFFF}U & \cellcolor[HTML]{FFFFFF}150 & \cellcolor[HTML]{FFFFFF}0.9 & \cellcolor[HTML]{FFFFFF}1.1 & \cellcolor[HTML]{FFFFFF}2.7 & \cellcolor[HTML]{FFFFFF} & \cellcolor[HTML]{FFFFFF}0.5 & \cellcolor[HTML]{FFFFFF}0.37 \\
\cellcolor[HTML]{FFFFFF}SkyHook (Helikites) & \cellcolor[HTML]{FFFFFF}B & \cellcolor[HTML]{FFFFFF}U & \cellcolor[HTML]{FFFFFF}2286 & \cellcolor[HTML]{FFFFFF}7.31 & \cellcolor[HTML]{FFFFFF}5.48 & \cellcolor[HTML]{FFFFFF} & \cellcolor[HTML]{FFFFFF}Fixed & \cellcolor[HTML]{FFFFFF}40 & \cellcolor[HTML]{FFFFFF}Tethered \\
\cellcolor[HTML]{FFFFFF}Zepellin-NT & \cellcolor[HTML]{FFFFFF}S & \cellcolor[HTML]{FFFFFF}M & \cellcolor[HTML]{FFFFFF}2600 & \cellcolor[HTML]{FFFFFF}75 & \cellcolor[HTML]{FFFFFF}19.5 & \cellcolor[HTML]{FFFFFF}8790 & \cellcolor[HTML]{FFFFFF}900 & \cellcolor[HTML]{FFFFFF}1900 & \cellcolor[HTML]{FFFFFF}24\\
\cellcolor[HTML]{FFFFFF}MD4-1000 (DHL) & \cellcolor[HTML]{FFFFFF}C & \cellcolor[HTML]{FFFFFF}U & \cellcolor[HTML]{FFFFFF}3000 & \cellcolor[HTML]{FFFFFF}1.03 & \cellcolor[HTML]{FFFFFF}1.03 & \cellcolor[HTML]{FFFFFF}2.9 & \cellcolor[HTML]{FFFFFF}20 & \cellcolor[HTML]{FFFFFF}1.2 & \cellcolor[HTML]{FFFFFF}0.8\\
\cellcolor[HTML]{FFFFFF}Skyship 600 (Charly) & \cellcolor[HTML]{FFFFFF}S & \cellcolor[HTML]{FFFFFF}M & \cellcolor[HTML]{FFFFFF}3050 & \cellcolor[HTML]{FFFFFF}59 & \cellcolor[HTML]{FFFFFF}15.2 & \cellcolor[HTML]{FFFFFF}3757 & \cellcolor[HTML]{FFFFFF}1019 & \cellcolor[HTML]{FFFFFF}2343 & \cellcolor[HTML]{FFFFFF}52\\
\cellcolor[HTML]{FFFFFF}Desert Star (Helikites) & \cellcolor[HTML]{FFFFFF}B & \cellcolor[HTML]{FFFFFF}U & \cellcolor[HTML]{FFFFFF}3352 & \cellcolor[HTML]{FFFFFF}10.05 & \cellcolor[HTML]{FFFFFF}6.7 & \cellcolor[HTML]{FFFFFF} & \cellcolor[HTML]{FFFFFF}Fixed & \cellcolor[HTML]{FFFFFF}100 & \cellcolor[HTML]{FFFFFF}Tethered\\
\cellcolor[HTML]{FFFFFF}MRI P2006T & \cellcolor[HTML]{FFFFFF}P & \cellcolor[HTML]{FFFFFF}M & \cellcolor[HTML]{FFFFFF}4200 & \cellcolor[HTML]{FFFFFF}8.7 & \cellcolor[HTML]{FFFFFF}11.4 & \cellcolor[HTML]{FFFFFF}850 & \cellcolor[HTML]{FFFFFF}926 & \cellcolor[HTML]{FFFFFF}380 & \cellcolor[HTML]{FFFFFF}6\\
\cellcolor[HTML]{FFFFFF}Protonex & \cellcolor[HTML]{FFFFFF}P & \cellcolor[HTML]{FFFFFF}U & \cellcolor[HTML]{FFFFFF}4250 & \cellcolor[HTML]{FFFFFF} & \cellcolor[HTML]{FFFFFF}8.2 & \cellcolor[HTML]{FFFFFF}50 & \cellcolor[HTML]{FFFFFF}600 & \cellcolor[HTML]{FFFFFF}25 & \cellcolor[HTML]{FFFFFF}9\\
\multicolumn{1}{l}{} & \multicolumn{1}{l}{} & \multicolumn{1}{l}{} & \multicolumn{1}{l}{} & \multicolumn{1}{l}{} & \multicolumn{1}{l}{} & \multicolumn{1}{l}{} & \multicolumn{1}{l}{} & \multicolumn{1}{l}{} \\ 
\cellcolor[HTML]{C0C0C0}\textbf{Medium Layer - MAP} & \multicolumn{1}{l}{\cellcolor[HTML]{C0C0C0}\textbf{}} & \multicolumn{1}{l}{\cellcolor[HTML]{C0C0C0}\textbf{}} & \multicolumn{1}{l}{\cellcolor[HTML]{C0C0C0}\textbf{}} & \multicolumn{1}{l}{\cellcolor[HTML]{C0C0C0}\textbf{}} & \multicolumn{1}{l}{\cellcolor[HTML]{C0C0C0}\textbf{}} & \multicolumn{1}{l}{\cellcolor[HTML]{C0C0C0}\textbf{}} & \multicolumn{1}{l}{\cellcolor[HTML]{C0C0C0}\textbf{}} & \multicolumn{1}{l}{\cellcolor[HTML]{C0C0C0}\textbf{}} & \multicolumn{1}{l}{\cellcolor[HTML]{C0C0C0}\textbf{}} \\
\cellcolor[HTML]{FFFFFF}Schiebel Camcopter S-100 & \cellcolor[HTML]{FFFFFF}P & \cellcolor[HTML]{FFFFFF}U & \cellcolor[HTML]{FFFFFF}5486 & \cellcolor[HTML]{FFFFFF}3.11 & \cellcolor[HTML]{FFFFFF}1.24 & \cellcolor[HTML]{FFFFFF}110 & \cellcolor[HTML]{FFFFFF}180 & \cellcolor[HTML]{FFFFFF}34 & \cellcolor[HTML]{FFFFFF}6\\
\cellcolor[HTML]{FFFFFF}ScanEagle & \cellcolor[HTML]{FFFFFF}P & \cellcolor[HTML]{FFFFFF}U & \cellcolor[HTML]{FFFFFF}5944 & \cellcolor[HTML]{FFFFFF}1.6 & \cellcolor[HTML]{FFFFFF}3.1 & \cellcolor[HTML]{FFFFFF}16 & \cellcolor[HTML]{FFFFFF} & \cellcolor[HTML]{FFFFFF}7.1 & \cellcolor[HTML]{FFFFFF}15\\
\cellcolor[HTML]{FFFFFF}Airlander 10 & \cellcolor[HTML]{FFFFFF}S & \cellcolor[HTML]{FFFFFF}M & \cellcolor[HTML]{FFFFFF}6100 & \cellcolor[HTML]{FFFFFF}92 & \cellcolor[HTML]{FFFFFF}43 & \cellcolor[HTML]{FFFFFF}20000 & \cellcolor[HTML]{FFFFFF} & \cellcolor[HTML]{FFFFFF}10000 & \cellcolor[HTML]{FFFFFF}504\\
\cellcolor[HTML]{FFFFFF}General Atomics Prowler II & \cellcolor[HTML]{FFFFFF}P & \cellcolor[HTML]{FFFFFF}U & \cellcolor[HTML]{FFFFFF}7600 & \cellcolor[HTML]{FFFFFF}5 & \cellcolor[HTML]{FFFFFF}10.75 & \cellcolor[HTML]{FFFFFF}250 & \cellcolor[HTML]{FFFFFF}2000 & \cellcolor[HTML]{FFFFFF}270 & \cellcolor[HTML]{FFFFFF}48\\
\cellcolor[HTML]{FFFFFF}FOTROS & \cellcolor[HTML]{FFFFFF}P & \cellcolor[HTML]{FFFFFF}U & \cellcolor[HTML]{FFFFFF}7600 & \cellcolor[HTML]{FFFFFF}6.2 & \cellcolor[HTML]{FFFFFF}17 & \cellcolor[HTML]{FFFFFF} & \cellcolor[HTML]{FFFFFF}2000 & \cellcolor[HTML]{FFFFFF} & \cellcolor[HTML]{FFFFFF}30\\
\cellcolor[HTML]{FFFFFF}{\color[HTML]{333333} EADS SDE Eagle 1} & \cellcolor[HTML]{FFFFFF}{\color[HTML]{333333} P} & \cellcolor[HTML]{FFFFFF}{\color[HTML]{333333} U} & \cellcolor[HTML]{FFFFFF}{\color[HTML]{333333} 7620} & \cellcolor[HTML]{FFFFFF}{\color[HTML]{333333} 9.3} & \cellcolor[HTML]{FFFFFF}{\color[HTML]{333333} 16.6} & \cellcolor[HTML]{FFFFFF}{\color[HTML]{333333} 1000} & \cellcolor[HTML]{FFFFFF}{\color[HTML]{333333} 1000} & \cellcolor[HTML]{FFFFFF}{\color[HTML]{333333} 250} & \cellcolor[HTML]{FFFFFF}{\color[HTML]{333333} 24}\\
\cellcolor[HTML]{FFFFFF}{\color[HTML]{333333} Solar Impulse 2} & \cellcolor[HTML]{FFFFFF}{\color[HTML]{333333} P} & \cellcolor[HTML]{FFFFFF}{\color[HTML]{333333} M} & \cellcolor[HTML]{FFFFFF}{\color[HTML]{333333} 8534} & \cellcolor[HTML]{FFFFFF}{\color[HTML]{333333} 22.4} & \cellcolor[HTML]{FFFFFF}{\color[HTML]{333333} 72} & \cellcolor[HTML]{FFFFFF}{\color[HTML]{333333} 2300} & \cellcolor[HTML]{FFFFFF}{\color[HTML]{333333} } & \cellcolor[HTML]{FFFFFF}{\color[HTML]{333333} 408} & \cellcolor[HTML]{FFFFFF}{\color[HTML]{333333} 117}\\
\cellcolor[HTML]{FFFFFF}{\color[HTML]{333333} MQ-1 Predator} & \cellcolor[HTML]{FFFFFF}{\color[HTML]{333333} P} & \cellcolor[HTML]{FFFFFF}{\color[HTML]{333333} U} & \cellcolor[HTML]{FFFFFF}{\color[HTML]{333333} 8839} & \cellcolor[HTML]{FFFFFF}{\color[HTML]{333333} 8.53} & \cellcolor[HTML]{FFFFFF}{\color[HTML]{333333} 17} & \cellcolor[HTML]{FFFFFF}{\color[HTML]{333333} 1233} & \cellcolor[HTML]{FFFFFF}{\color[HTML]{333333} 400} & \cellcolor[HTML]{FFFFFF}{\color[HTML]{333333} 487} & \cellcolor[HTML]{FFFFFF}{\color[HTML]{333333} 24}\\
\cellcolor[HTML]{FFFFFF}{\color[HTML]{333333} Anka - A} & \cellcolor[HTML]{FFFFFF}{\color[HTML]{333333} P} & \cellcolor[HTML]{FFFFFF}{\color[HTML]{333333} U} & \cellcolor[HTML]{FFFFFF}{\color[HTML]{333333} 9144} & \cellcolor[HTML]{FFFFFF}{\color[HTML]{333333} 8} & \cellcolor[HTML]{FFFFFF}{\color[HTML]{333333} 17.3} & \cellcolor[HTML]{FFFFFF}{\color[HTML]{333333} 1400} & \cellcolor[HTML]{FFFFFF}{\color[HTML]{333333} 4896} & \cellcolor[HTML]{FFFFFF}{\color[HTML]{333333} 200} & \cellcolor[HTML]{FFFFFF}{\color[HTML]{333333} 24}\\
\cellcolor[HTML]{FFFFFF}{\color[HTML]{333333} Silver Arrow Sniper} & \cellcolor[HTML]{FFFFFF}{\color[HTML]{333333} P} & \cellcolor[HTML]{FFFFFF}{\color[HTML]{333333} U} & \cellcolor[HTML]{FFFFFF}{\color[HTML]{333333} 9145} & \cellcolor[HTML]{FFFFFF}{\color[HTML]{333333} 9.4} & \cellcolor[HTML]{FFFFFF}{\color[HTML]{333333} 18} & \cellcolor[HTML]{FFFFFF}{\color[HTML]{333333} 1250} & \cellcolor[HTML]{FFFFFF}{\color[HTML]{333333} 200} & \cellcolor[HTML]{FFFFFF}{\color[HTML]{333333} 400} & \cellcolor[HTML]{FFFFFF}{\color[HTML]{333333} 26}\\
\multicolumn{1}{l}{} & \multicolumn{1}{l}{} & \multicolumn{1}{l}{} & \multicolumn{1}{l}{} & \multicolumn{1}{l}{} & \multicolumn{1}{l}{} & \multicolumn{1}{l}{} & \multicolumn{1}{l}{} & \multicolumn{1}{l}{} \\ 
\cellcolor[HTML]{C0C0C0}\textbf{Higher Layer - HAP} & \multicolumn{1}{l}{\cellcolor[HTML]{C0C0C0}} & \multicolumn{1}{l}{\cellcolor[HTML]{C0C0C0}} & \multicolumn{1}{l}{\cellcolor[HTML]{C0C0C0}} & \multicolumn{1}{l}{\cellcolor[HTML]{C0C0C0}} & \multicolumn{1}{l}{\cellcolor[HTML]{C0C0C0}} & \multicolumn{1}{l}{\cellcolor[HTML]{C0C0C0}} & \multicolumn{1}{l}{\cellcolor[HTML]{C0C0C0}} & \multicolumn{1}{l}{\cellcolor[HTML]{C0C0C0}} & \multicolumn{1}{l}{\cellcolor[HTML]{C0C0C0}}\\
\cellcolor[HTML]{FFFFFF}{\color[HTML]{333333} IAI Heron} & \cellcolor[HTML]{FFFFFF}{\color[HTML]{333333} P} & \cellcolor[HTML]{FFFFFF}{\color[HTML]{333333} U} & \cellcolor[HTML]{FFFFFF}{\color[HTML]{333333} 10000} & \cellcolor[HTML]{FFFFFF}{\color[HTML]{333333} 8.5} & \cellcolor[HTML]{FFFFFF}{\color[HTML]{333333} 16.6} & \cellcolor[HTML]{FFFFFF}{\color[HTML]{333333} 900} & \cellcolor[HTML]{FFFFFF}{\color[HTML]{333333} 350} & \cellcolor[HTML]{FFFFFF}{\color[HTML]{333333} 250} & \cellcolor[HTML]{FFFFFF}{\color[HTML]{333333} 52}\\
\cellcolor[HTML]{FFFFFF}{\color[HTML]{333333} Predator B (MQ-9B)} & \cellcolor[HTML]{FFFFFF}{\color[HTML]{333333} P} & \cellcolor[HTML]{FFFFFF}{\color[HTML]{333333} U} & \cellcolor[HTML]{FFFFFF}{\color[HTML]{333333} 15000} & \cellcolor[HTML]{FFFFFF}{\color[HTML]{333333} 11} & \cellcolor[HTML]{FFFFFF}{\color[HTML]{333333} 20} & \cellcolor[HTML]{FFFFFF}{\color[HTML]{333333} 2223} & \cellcolor[HTML]{FFFFFF}{\color[HTML]{333333} 1852} & \cellcolor[HTML]{FFFFFF}{\color[HTML]{333333} 1700} & \cellcolor[HTML]{FFFFFF}{\color[HTML]{333333} 14}\\
\cellcolor[HTML]{FFFFFF}{\color[HTML]{333333} G520 Strto 1} & \cellcolor[HTML]{FFFFFF}{\color[HTML]{333333} P} & \cellcolor[HTML]{FFFFFF}{\color[HTML]{333333} M} & \cellcolor[HTML]{FFFFFF}{\color[HTML]{333333} 16000} & \cellcolor[HTML]{FFFFFF}{\color[HTML]{333333} 13.82} & \cellcolor[HTML]{FFFFFF}{\color[HTML]{333333} 33} & \cellcolor[HTML]{FFFFFF}{\color[HTML]{333333} 3300} & \cellcolor[HTML]{FFFFFF}{\color[HTML]{333333} 3670} & \cellcolor[HTML]{FFFFFF}{\color[HTML]{333333} 1400} & \cellcolor[HTML]{FFFFFF}{\color[HTML]{333333} 8} \\
\cellcolor[HTML]{FFFFFF}{\color[HTML]{333333} Northrop Grumman Global Hawk} & \cellcolor[HTML]{FFFFFF}{\color[HTML]{333333} P} & \cellcolor[HTML]{FFFFFF}{\color[HTML]{333333} U} & \cellcolor[HTML]{FFFFFF}{\color[HTML]{333333} 18000} & \cellcolor[HTML]{FFFFFF}{\color[HTML]{333333} 14.5} & \cellcolor[HTML]{FFFFFF}{\color[HTML]{333333} 39.9} & \cellcolor[HTML]{FFFFFF}{\color[HTML]{333333} 6781} & \cellcolor[HTML]{FFFFFF}{\color[HTML]{333333} 22779} & \cellcolor[HTML]{FFFFFF}{\color[HTML]{333333} 1360} & \cellcolor[HTML]{FFFFFF}{\color[HTML]{333333} 32} \\
\cellcolor[HTML]{FFFFFF}{\color[HTML]{333333} Zephyr 6} & \cellcolor[HTML]{FFFFFF}{\color[HTML]{333333} P} & \cellcolor[HTML]{FFFFFF}{\color[HTML]{333333} U} & \cellcolor[HTML]{FFFFFF}{\color[HTML]{333333} 18288} & \cellcolor[HTML]{FFFFFF}{\color[HTML]{333333} } & \cellcolor[HTML]{FFFFFF}{\color[HTML]{333333} 18} & \cellcolor[HTML]{FFFFFF}{\color[HTML]{333333} 30} & \cellcolor[HTML]{FFFFFF}{\color[HTML]{333333} } & \cellcolor[HTML]{FFFFFF}{\color[HTML]{333333} 2.27} & \cellcolor[HTML]{FFFFFF}{\color[HTML]{333333} 30}\\
\cellcolor[HTML]{FFFFFF}{\color[HTML]{333333} Aurora Flight Sciences Perseus} & \cellcolor[HTML]{FFFFFF}{\color[HTML]{333333} P} & \cellcolor[HTML]{FFFFFF}{\color[HTML]{333333} U} & \cellcolor[HTML]{FFFFFF}{\color[HTML]{333333} 19812} & \cellcolor[HTML]{FFFFFF}{\color[HTML]{333333} 7.62} & \cellcolor[HTML]{FFFFFF}{\color[HTML]{333333} 21.79} & \cellcolor[HTML]{FFFFFF}{\color[HTML]{333333} 1936} & \cellcolor[HTML]{FFFFFF}{\color[HTML]{333333} } & \cellcolor[HTML]{FFFFFF}{\color[HTML]{333333} 99.79} & \cellcolor[HTML]{FFFFFF}{\color[HTML]{333333} 24} \\
\cellcolor[HTML]{FFFFFF}{\color[HTML]{333333} Stratobus} & \cellcolor[HTML]{FFFFFF}{\color[HTML]{333333} S} & \cellcolor[HTML]{FFFFFF}{\color[HTML]{333333} U} & \cellcolor[HTML]{FFFFFF}{\color[HTML]{333333} 20000} & \cellcolor[HTML]{FFFFFF}{\color[HTML]{333333} 100} & \cellcolor[HTML]{FFFFFF}{\color[HTML]{333333} 30} & \cellcolor[HTML]{FFFFFF}{\color[HTML]{333333} } & \cellcolor[HTML]{FFFFFF}{\color[HTML]{333333} } & \cellcolor[HTML]{FFFFFF}{\color[HTML]{333333} 250} & \cellcolor[HTML]{FFFFFF}{\color[HTML]{333333} 5 years} \\
\cellcolor[HTML]{FFFFFF}{\color[HTML]{333333} M-55 Geophysica} & \cellcolor[HTML]{FFFFFF}{\color[HTML]{333333} P} & \cellcolor[HTML]{FFFFFF}{\color[HTML]{333333} M} & \cellcolor[HTML]{FFFFFF}{\color[HTML]{333333} 21000} & \cellcolor[HTML]{FFFFFF}{\color[HTML]{333333} 22.86} & \cellcolor[HTML]{FFFFFF}{\color[HTML]{333333} 37.46} & \cellcolor[HTML]{FFFFFF}{\color[HTML]{333333} 13995} & \cellcolor[HTML]{FFFFFF}{\color[HTML]{333333} 4965} & \cellcolor[HTML]{FFFFFF}{\color[HTML]{333333} 7000} & \cellcolor[HTML]{FFFFFF}{\color[HTML]{333333} 6.5} \\
\cellcolor[HTML]{FFFFFF}{\color[HTML]{333333} ISIS (Integrated Sensor is Structure)} & \cellcolor[HTML]{FFFFFF}{\color[HTML]{333333} S} & \cellcolor[HTML]{FFFFFF}{\color[HTML]{333333} U} & \cellcolor[HTML]{FFFFFF}{\color[HTML]{333333} 21500} & \cellcolor[HTML]{FFFFFF}{\color[HTML]{333333} 137.16} & \cellcolor[HTML]{FFFFFF}{\color[HTML]{333333} } & \cellcolor[HTML]{FFFFFF}{\color[HTML]{333333} } & \cellcolor[HTML]{FFFFFF}{\color[HTML]{333333} } & \cellcolor[HTML]{FFFFFF}{\color[HTML]{333333} 2700} & \cellcolor[HTML]{FFFFFF}{\color[HTML]{333333} 10 years}\\
\hline
\end{tabular}}
\end{table*}

\subsection{Functionalities of Layers in Proposed Architecture:}
The general system architecture of our proposed \ac{NFP} has been divided into three layers of operation. The \ac{HL} ranging between 15km-25km, the \ac{ML} ranging between 5km-15km and the \ac{LL} which is up to 5km height. For each layer, the operating conditions can vary in terms of wind speed, humidity and temperature. This means that certain type of platforms can be operational at a particular layer, that can cope with the flying conditions as well as provide their services which can range from surveillance, broadband connectivity to backhaul communications as described below.

\subsubsection{\ac{HAP}} is a stratospheric platform capable of delivering a wide range of services such as mobile cellular communications, broadband wireless access as well as search and rescue services. A \ac{HAP} is expected to operate at the \ac{HL} providing \ac{LOS} connectivity over a wide geographical area (30km radius). Such platforms can either be planes or airships, manned or unmanned that can carry a payload of a few kilograms to a few tones and can stay airborne from a few hours to a few years depending on their type, size, power constraints and fuel capacity. A \ac{HAP} can provide ubiquitous coverage providing backhaul and control/fleet coordination services for other aerial platforms at lower layers. A fleet of HAPs can be used to provide extended communication coverage and redundancy if necessary, while inter-platform communications can be established employing \ac{FSO}.     
\subsubsection{\ac{MAP}} are aerial platforms operating at the \ac{ML} that can be used as a relay between a \ac{HAP} and a \ac{LAP}. Depending on the operation scenario, current available \ac{MAP} in the market are mostly \ac{UAV} with long endurance capabilities as well as manned aerial vehicles. \ac{UAV} platforms can stay airborne for several hours and are usually destined for military missions. \ac{MAP} coverage area is expected to be of up to 5km radius. 
\subsubsection{\ac{LAP}} such as tethered balloons, drones, operate	at the \ac{LL}. Like \ac{HAP} and \ac{MAP}, \ac{LAP} exhibit common features such  as \ac{LOS} communications with favorable radio conditions, while they have the ability	to rapidly deploy a fleet of \ac{LAP} with modular communication payload capabilities. \ac{LAP} are currently in high demand for public usage and the current market is driving the drone industry in delivering newer and smarter platforms, with higher endurance and greater range, more agile and efficient aerial vehicles. \ac{LAP} are optimally distributed to offer capacity and expand coverage via resource and interference management. 

Table 1 lists a number of aerial platforms that can be employed to carry out a mission on a particular layer. In \ac{LL}, \ac{LAP} are expected to feature a relatively small size, limited payload and endurance capabilities in minutes or hours while offering rapid deployment operating below 5km height. \ac{LAP} are expected to operate in organised / scheduled shifts in order to ensure continuity of service. \ac{MAP} are expected to be able to cope with a heavier payload in order to maintain their role as network relays between \ac{LL} and \ac{ML} featuring more functionalities than \ac{LAP}. They are capable of staying aloft for a greater duration of time (hours / days) in order to support the operation of \ac{LAP}. \ac{MAP} are expect to be operational between 5-15km height subject to the local aviation regulations of operation. Finally, \ac{HAP} operating on \ac{HL} will be of much bigger size operating at a height around 17-20km height. Such platforms will be carrying a much heavier payload that would support \ac{RF} and \ac{FSO} communications. It should be able to stay operational for days/months without require refueling. A possible scenario is to use Protonex as \ac{LAP} that can stay aloft up to 9 hours while the Airlander10 can operate as \ac{MAP} providing services at about 10km height. Finally, a \ac{HAP} such as ISIS can stay aloft for a great duration of time providing connectivity to the \ac{MAP}.  

\section{Optimisation of Airbone \ac{SON}}
%\subsection{Overview of the network}
%In an airborne surveillance system with reaction to the intruder, the action begins with the detection of the intruder. Localization and tracking of the intruder are the next steps that are coordinated by the centralized modules using machine learning techniques. In \cite{pandey2011scene_Localization} the authors discuss these techniques in detail. The final steps are jamming intruder's signal and spoofing it \cite{tippenhauer2011_spoofingReq}. In an airborne system, for these steps the jammer(s) should to be close to the intruder \cite{kerns2014_UAVcontrol}. Therefore, some \ac{UAV}s (Attacking \ac{UAV}s) will leave their position and the network should be able to reconfigure itself in a way to seamlessly handover the users of attacking \ac{UAV}s.

%dynamic network -- self-organizing in different layers
%\subsection{Airborne \ac{SON}}

In our proposed architecture the position of \ac{LAP}s in the \ac{LL} is defined centrally and the \ac{NFP} has the ability to re-organise the \ac{LL} to achieve its target, which can be capturing as many \ac{UE}s, maximizing the achievable rate, and/or fairness among \ac{UE}s. However, optimum placement is an NP-hard problem. 

\ac{NFP}-\ac{LL} placement to capture maximum \ac{UE}s can be formulated as:
\begin{align*} 
\text max & \sum_{n=1}^N\sum_{u=1}^U d_{n,u} & \\ 
\text{subject to} &\sum_{u=1}^U d_{n,u}\leq 1, &\forall n\in\{1,\dots,N\} 
\\
& d_{n,u} = \left\{\begin{matrix}
1 & RSS_{n,u}>RSS_{-n,u}\\ 
0 & otherwise
\end{matrix}\right. & \forall u\in\{1,\dots,U\},
\end{align*}
where $N$ is the number of \ac{LAP}s in \ac{NFP}-\ac{LL}, $U$ is the number of \ac{UE}s, and $d_{n,u}$ is $1$ if the $u^{th}$  \ac{UE} is served by \ac{LAP} $n$, otherwise it will be zero. $RSS_{n,u}$ denotes the received signal strength of \ac{UE} $u$ from \ac{LAP} $n$. $RSS_{-n,u}$ denotes the strengths of received signal from other nodes of \ac{NFP} and the Macrocells, if exists.

The above mixed integer nonlinear problem can be linearized using the technique presented in \cite{ahmadi2013framework} and commercial packages like CPLEX can solve the mixed integer linear problem. Problems with limited number of \ac{UE}s, and \ac{LAP}s serving as hotspots can be solved with exhaustive search. In this work we use this technique. However, in more complex scenarios even the mixed integer linear problem will be too complex. Therefore, metaheuristic algorithms can be used to find an efficient sub-optimum solution \cite{ahmadi2012evolutionary}. We introduce some of these metaheuristic methods.

\subsection{Resource allocation using metaheuristics:}
Popular metaheuristics in radio resource allocation are \ac{GA}, \ac{ACO}, and \ac{PSO}.

\subsubsection{\ac{GA}} code the solution into a chromosome (also known as individual) and evaluates the optimality of the solution with a function called fitness function \cite{Ahmadi_GA09}. In this work the fitness function is the sum captured \ac{UE}s, and the chromosome consists of the position of \ac{UAV}s. Crossover and mutation are tools of \ac{GA} for improving the solutions. Crossover combines fragments of two chromosomes and creates a new chromosome which is another solution in the feasible area. Mutation is the tool which is hired for overcoming the local optimums; mutation changes a gene randomly with the hope of reaching better solutions. The algorithm starts with a randomly generated population of chromosomes and at each iteration it creates a generation of offspring using crossover and mutation. The fittest of parent and offspring generations are kept for the next iteration and the rest are discarded. This way \ac{GA} keep the population size constant. The algorithm stops after the maximum number of iterations is reached.

\subsubsection{\ac{ACO}} has been used in radio resource management for network planning and spectrum allocation \cite{Ahmadi_ACO10}. The main idea is taken from the way ants find the shortest path to the food using pheromone. The shorter and more popular paths have higher density of pheromone while longer paths will lose pheromone due to the evaporation. In this model each solution will be coded as a path, and at each iteration each ant chooses a path with a probability that is proportional to the amount of pheromone. After a certain number of iterations the most popular will be selected, i.e. the algorithm converges. 
\subsubsection{\ac{PSO}} models the social behavior of a group of birds \cite{PSO_kulkarni11}. It consists of a swarm of candidate solutions called particles, which explore an n-dimensional hyperspace in search of the global solution. Compared to \ac{GA} and \ac{ACO} modeling the mixed integer problems is more straightforward in \ac{PSO}.

\subsection{Decentralized decision making:}
In the proposed architecture we use centralized decision making to define the position of \ac{LAP} in \ac{LL}. However, the architecture has the potential to integrate a distributed decision making mechanism for positioning the \ac{LAP}. Game theory is the most popular techniques used for designing and analyzing distributed decision making approaches \cite{mackenzie2006game}. Different classes of game theoretic approaches can be used to model competition, cooperation, and coalition between different players, \ac{LAP}s. Game theoretic analysis of the system tells that if there exists a path that leads the competition/cooperation of independent decision makers to an equilibrium. In other words, it enables the system designers to avoid objective functions that lead the system to instability.   

%\subsubsection{Learning techniques in airborne \ac{SON}} 

In an airborne network settings,  parameters of the network will change much faster than a terrestrial network. Therefore, learning from past experiences is extremely important in both centralized and decentralized decision making mechanisms for airborne \ac{SON} \cite{Gavrilovska_learning13}. In decentralized systems, learning techniques help reaching an equilibrium in fewer iterations.

%However, learning and reasoning techniques  more crucial in decentralized decision making

%The \ac{SON} must be capable to learn from past experiences \cite{Gavrilovska_learning13}

%The main elements of self-configuration for a \ac{BS} is IP address self configuration, neighbor cell list self-configuration, and radio access parameters self-configuration.     

%However, a centralized system can also self-organise (re-organise) to improve its functionality. 

%\subsection{resource allocation problem in \ac{SON}}

%\subsubsection{How many and which UAVs should attack?}

%\subsubsection{Replacement for coverage/capacity}

%spoofing \cite{tippenhauer2011_spoofingReq}
%\cite{kerns2014_UAVcontrol}

\begin{figure}[!ht]
	\begin{center}
		\includegraphics[width=0.9\columnwidth]{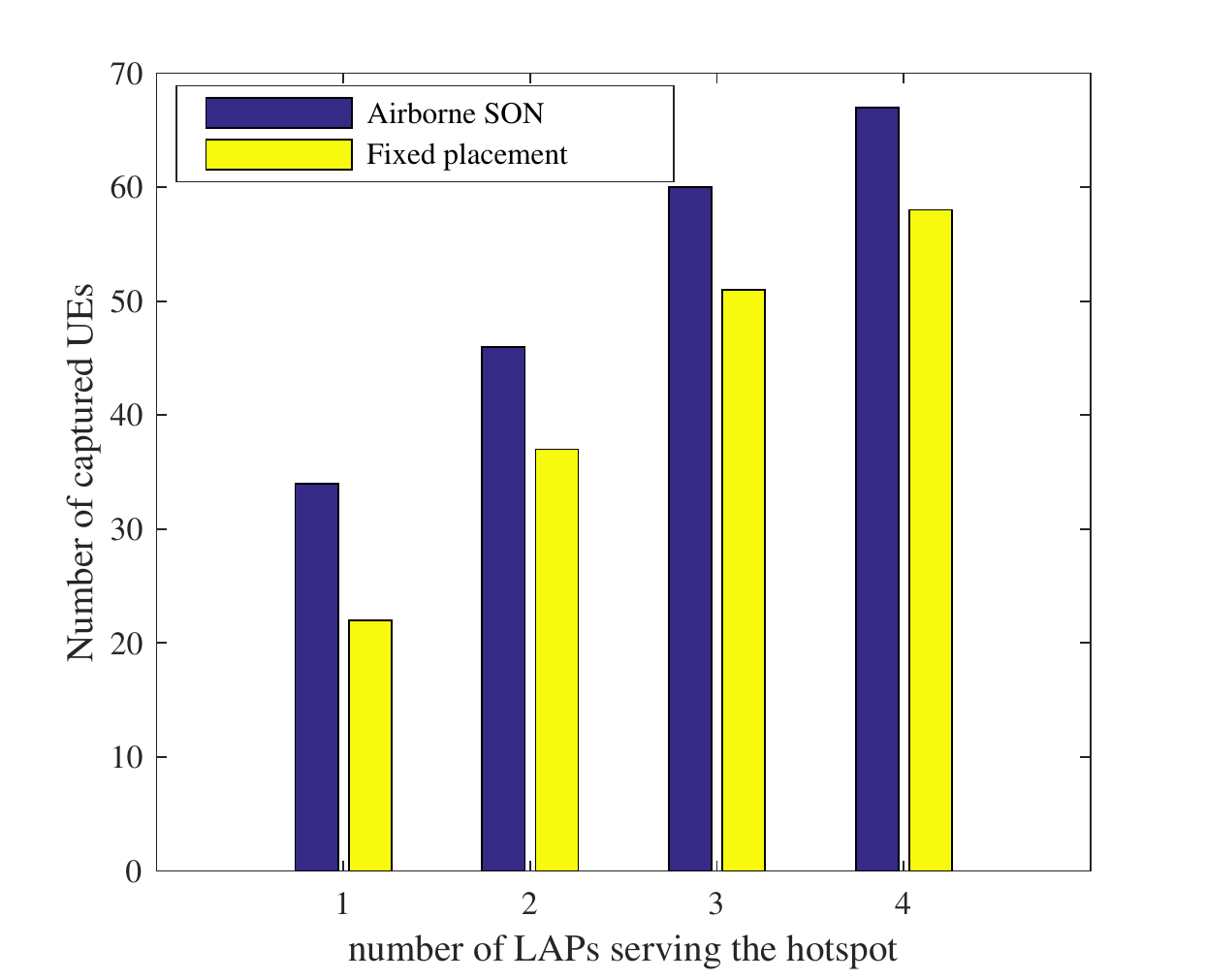}\end{center}
	
    \caption{Number of UEs captured by airborne SON compared to the fixed placement of LAPs in NFP-LL. The network has 150 UEs where a third of them are in the demand hotspot. Demand hotspot has a radius of 250 m (e.g. outdoor concert area).}
	\label{illustrative3}
\end{figure}

\section{Simulation results and discussions}
% \begin{figure}
% 	% Requires \usepackage{graphicx}
% 	\begin{center}
% 	\includegraphics[width=0.8\columnwidth]{./Figure/res2}
% 	\end{center}
% 	\caption{sample scenario}
% 	\label{illustrative2}
% \end{figure}

We modeled an \ac{NFP} where the \ac{LL} provides service to a demand hotspot. The \ac{UE}s in the system are served by the macrocell and the \ac{NFP}-\ac{LL} assists the macrocell by capturing \ac{UE}s. The \ac{NFP} optimizes its \ac{LL} for a given number of \ac{LAP}s. Due to weather conditions, battery failure or surveillance duty the system may lose an \ac{LAP}. In this work, we compared an airborne \ac{SON} system that reorganizes itself with an \ac{NFP} with fixed \ac{LL} placement. Our results in Fig.~\ref{illustrative3} shows that the airborne \ac{SON} outperforms the fixed placement. Cross-layer optimisation approaches are required to optimise the placement or positioning of \ac{NFP}s across the layers in proposed airborne \ac{SON}. 

\section{Challenges and future directions}
Although airborne systems have attracted industry and academia's attention in the last couple of years, there still exists several challenges and open research directions. Following are some of the important challenges for future airborne \ac{SON}:

\subsection{Standardization}
Airborne cellular networks are yet to be standardized. The existing networking standards cannot fully address the challenges of airborne networks and proper standards for airborne communication and networking is required.

\subsection{Surveillance}
Airborne cellular network would offer complementary connectivity services to expand the coverage or inject the capacity under some unknown situations, therefore their successful operation would depend on advanced surveillance mechanisms to detect amateur flying platforms and combat to avoid any further disruption in cellular services. 

\subsection{Information security}
Secure transmission of information over wireless links is still a challenge for future Internet architecture. In our proposed system, \ac{NFP}s in all layers have the ability to move and complement the existing network, therefore, the dynamic nature of the airborne cellular network could be an additional challenge to ensure secure coverage expansion or capacity enhancement. 

\subsection{Ethics and privacy}
\ac{NFP}s and swarm of \ac{NFP}s may face two-fold challenges in order to comply with regulatory issues related to privacy and ethics. \ac{NFP}s should be able to protect the privacy of the connected users while following the flying ethics as per regulations and avoiding no-flying zone.  
 
\subsection{Testbed and verification} 
Various projects in Europe and United States study and test the performance of future Internet and connectivity architecture, resource allocation techniques, waveforms, and integration of future technologies using advanced testbeds \cite{fireProject}. To the best of our knowledge, none of the existing testbed validation and experimentation provide an environment for testing the proposed airborne \ac{SON}.

\section{Conclusions}
In this paper we have presented a novel layered architecture where NFPs, of various types and flying in low/medium/high layers in a swarm of flying platforms, are considered as an integrated part of the future cellular networks to inject additional capacity and expand the coverage for exceptional scenarios and hard-to-reach areas. In our proposed architecture the position of LAPs in the LL is defined centrally and the NFP has the ability to re-organise the LL to achieve its target, which can be capturing as many UEs, maximizing the achievable rate, and/or fairness among UEs. To evaluate the proposed architecture, we compared an airborne SON system that reorganizes itself with an NFP with fixed LL placement. Our results show that the airborne SON outperforms the fixed placement.

%\appendices

%\section{}

%\section*{Acknowledgment}


\ifCLASSOPTIONcaptionsoff
  \newpage
\fi



%\bibliographystyle{IEEEtran}
%\vspace{-0.3cm}
\balance
\begin{acronym} 
\acro{5G}{Fifth Generation}
\acro{ACO}{Ant Colony Optimization}
\acro{BB}{Base Band}
\acro{BBU}{Base Band Unit}
\acro{BER}{Bit Error Rate}
\acro{BS}{Base Station}
\acro{BW}{bandwidth}
\acro{C-RAN}{Cloud Radio Access Networks}
\acro{CAPEX}{Capital Expenditure}
\acro{CoMP}{Coordinated Multipoint}
\acro{DAC}{Digital-to-Analog Converter}
\acro{DAS}{Distributed Antenna Systems}
\acro{DBA}{Dynamic Bandwidth Allocation}
\acro{DL}{Downlink}
\acro{FBMC}{Filterbank Multicarrier}
\acro{FEC}{Forward Error Correction}
\acro{FFR}{Fractional Frequency Reuse}
\acro{FSO}{Free Space Optics}
\acro{GA}{Genetic Algorithms}
\acro{HAP}{High Altitude Platform}
\acro{HL}{Higher Layer}
\acro{HARQ}{Hybrid-Automatic Repeat Request}
\acro{IoT}{Internet of Things}
\acro{LAN}{Local Area Network}
\acro{LAP}{Low Altitude Platform}
\acro{LL}{Lower Layer}
\acro{LOS}{Line of Sight}
\acro{LTE}{Long Term Evolution}
\acro{LTE-A}{Long Term Evolution Advanced}
\acro{MAC}{Medium Access Control}
\acro{MAP}{Medium Altitude Platform}
\acro{ML}{Medium Layer}
\acro{MME}{Mobility Management Entity}
\acro{mmWave}{millimeter Wave}
\acro{MIMO}{Multiple Input Multiple Output}
\acro{NFP}{Network Flying Platform}
\acro{NFPs}{Network Flying Platforms}
\acro{OFDM}{Orthogonal Frequency Division Multiplexing}
\acro{PAM}{Pulse Amplitude Modulation}
\acro{PAPR}{Peak-to-Average Power Ratio}
\acro{PGW}{Packet Gateway}
\acro{PHY}{physical layer}
\acro{PSO}{Particle Swarm Optimization}
\acro{QAM}{Quadrature Amplitude Modulation}
\acro{QoE}{Quality of Experience}
\acro{QoS}{Quality of Service}
\acro{QPSK}{Quadrature Phase Shift Keying}
\acro{RF}{Radio Frequency}
\acro{RN}{Remote Node}
\acro{RRH}{Remote Radio Head}
\acro{RRC}{Radio Resource Control}
\acro{RRU}{Remote Radio Unit}
\acro{SCBS}{Small Cell Base Station}
\acro{SDN}{Software Defined Network}
\acro{SNR}{Signal-to-Noise Ratio}
\acro{SON}{Self-organising Network}
\acro{TDD}{Time Division Duplex}
\acro{TD-LTE}{Time Division LTE}
\acro{TDM}{Time Division Multiplexing}
\acro{TDMA}{Time Division Multiple Access}
\acro{UE}{User Equipment}
\acro{UAV}{Unmanned Aerial Vehicle}

\end{acronym}

% Usage in text:
% \ac{BBU} for singular
% \acp{BBU} for plural
\bibliographystyle{IEEEtran}
\bibliography{UAV_ref}

\begin{thebibliography}{10}
\providecommand{\url}[1]{#1}
\csname url@samestyle\endcsname
\providecommand{\newblock}{\relax}
\providecommand{\bibinfo}[2]{#2}
\providecommand{\BIBentrySTDinterwordspacing}{\spaceskip=0pt\relax}
\providecommand{\BIBentryALTinterwordstretchfactor}{4}
\providecommand{\BIBentryALTinterwordspacing}{\spaceskip=\fontdimen2\font plus
\BIBentryALTinterwordstretchfactor\fontdimen3\font minus
  \fontdimen4\font\relax}
\providecommand{\BIBforeignlanguage}[2]{{%
\expandafter\ifx\csname l@#1\endcsname\relax
\typeout{** WARNING: IEEEtran.bst: No hyphenation pattern has been}%
\typeout{** loaded for the language `#1'. Using the pattern for}%
\typeout{** the default language instead.}%
\else
\language=\csname l@#1\endcsname
\fi
#2}}
\providecommand{\BIBdecl}{\relax}
\BIBdecl

\bibitem{6815890}
A.~Osseiran, F.~Boccardi, V.~Braun, K.~Kusume, P.~Marsch, M.~Maternia,
  O.~Queseth, M.~Schellmann, H.~Schotten, H.~Taoka, H.~Tullberg, M.~A.
  Uusitalo, B.~Timus, and M.~Fallgren, ``Scenarios for 5g mobile and wireless
  communications: the vision of the metis project,'' \emph{IEEE Communications
  Magazine}, vol.~52, no.~5, pp. 26--35, May 2014.

\bibitem{c2b90718cf1640aa91ce30953aae3b57}
D.~Grace, K.~Katzis, D.~Pearce, and P.~Mitchell, ``Low-latency mac-layer
  handoff for a high altitude platform delivering broadband communications,''
  \emph{URSI Radio Science Bulletin}, no. 332, pp. 39--49, 3 2010.

\bibitem{317e1315569e447aa94e59d77455937a}
K.~Katzis and D.~Grace, ``Inter-high altitude platform handoff for
  communications systems with directional antennas,'' \emph{URSI Radio Science
  Bulletin Special Issue on HAPs}, 6 2010.

\bibitem{ROHDE20131893}
\BIBentryALTinterwordspacing
S.~Rohde, M.~Putzke, and C.~Wietfeld, ``Ad hoc self-healing of ofdma networks
  using uav-based relays,'' \emph{Ad Hoc Networks}, vol.~11, no.~7, pp. 1893 --
  1906, 2013, theory, Algorithms and Applications of Wireless Networked
  Robotics Recent Advances in Vehicular Communications and Networking.
  [Online]. Available:
  \url{http://www.sciencedirect.com/science/article/pii/S157087051200131X}
\BIBentrySTDinterwordspacing

\bibitem{7122575}
X.~Li, D.~Guo, H.~Yin, and G.~Wei, ``Drone-assisted public safety wireless
  broadband network,'' in \emph{2015 IEEE Wireless Communications and
  Networking Conference Workshops (WCNCW)}, March 2015, pp. 323--328.

\bibitem{6863654}
A.~Al-Hourani, S.~Kandeepan, and S.~Lardner, ``Optimal lap altitude for maximum
  coverage,'' \emph{IEEE Wireless Communications Letters}, vol.~3, no.~6, pp.
  569--572, Dec 2014.

\bibitem{7510820}
R.~I. Bor-Yaliniz, A.~El-Keyi, and H.~Yanikomeroglu, ``Efficient 3-d placement
  of an aerial base station in next generation cellular networks,'' in
  \emph{2016 IEEE International Conference on Communications (ICC)}, May 2016,
  pp. 1--5.

\bibitem{aliu2013_SONsurvey}
O.~G. Aliu, A.~Imran, M.~A. Imran, and B.~Evans, ``A survey of self
  organisation in future cellular networks,'' \emph{IEEE Communications Surveys
  \& Tutorials}, vol.~15, no.~1, pp. 336--361, 2013.

\bibitem{icc2017wk}
E.~Kalantari, M.~Z. Shakir, H.~Yanikomeroglu, and A.~Yongacoglu,
  ``Backhaul-aware robust {3D} drone placement in {5G+} wireless networks,'' in
  \emph{Flexible Network Workshop (FlexNet), 2017 IEEE International Conference
  on Communications}, 2017.

\bibitem{ahmadi2013framework}
H.~Ahmadi, Y.~H. Chew, and C.~C. Chai, ``A framework to optimize the ofdm-based
  multiuser dynamic spectrum access networks with frequency reuse,''
  \emph{Wireless personal communications}, pp. 1--18, 2013.

\bibitem{ahmadi2012evolutionary}
H.~Ahmadi and Y.~Chew, ``Evolutionary algorithms for orthogonal frequency
  division multiplexing-based dynamic spectrum access systems,'' \emph{Computer
  Networks}, vol.~56, no.~14, pp. 3206--3218, 2012.

\bibitem{Ahmadi_GA09}
H.~Ahmadi and Y.~H. Chew, ``Adaptive subcarrier-and-bit allocation in
  multiclass multiuser ofdm systems using genetic algorithm,'' in \emph{2009
  IEEE 20th International Symposium on Personal, Indoor and Mobile Radio
  Communications}, Sept 2009, pp. 1883--1887.

\bibitem{Ahmadi_ACO10}
------, ``Subcarrier-and-bit allocation in multiclass multiuser single-cell
  ofdma systems using an ant colony optimization based evolutionary
  algorithm,'' in \emph{2010 IEEE Wireless Communication and Networking
  Conference}, April 2010, pp. 1--5.

\bibitem{PSO_kulkarni11}
R.~V. Kulkarni and G.~K. Venayagamoorthy, ``Particle swarm optimization in
  wireless-sensor networks: A brief survey,'' \emph{IEEE Transactions on
  Systems, Man, and Cybernetics, Part C (Applications and Reviews)}, vol.~41,
  no.~2, pp. 262--267, March 2011.

\bibitem{mackenzie2006game}
A.~B. MacKenzie and L.~A. DaSilva, ``Game theory for wireless engineers,''
  \emph{Synthesis Lectures on Communications}, vol.~1, no.~1, pp. 1--86, 2006.

\bibitem{Gavrilovska_learning13}
L.~Gavrilovska, V.~Atanasovski, I.~Macaluso, and L.~A. DaSilva, ``Learning and
  reasoning in cognitive radio networks,'' \emph{IEEE Communications Surveys
  Tutorials}, vol.~15, no.~4, pp. 1761--1777, Fourth 2013.

\bibitem{fireProject}
``Fire project,'' \url{https://www.ict-fire.eu}.

\end{thebibliography}


% \end{thebibliography}



%\newpage



%\vfill


\end{document}